\documentstyle[amssymb, preprint,aps]{revtex}
\begin{document}
\draft
\author{Yuri P. Malakyan}
\address{Institute for Physical Research, 378410, Ashtarak-2, Armenia \\
E-mail: yumal@ipr.sci.am\\
(December 10, 2001)}
\title{{\bf Temporal Oscillations of Nonlinear Faraday Rotation in Coherently
Driven Media}}
\maketitle

\begin{abstract}
New phenomenon of temporal oscillations of nonlinear Faraday rotation in a
driven four-level system is predicted. We show that in this system with one
upper level, under the conditions of electromagnetically induced
transparency created by a strong coupling field, the polarization rotation
of weak probe light exhibits slowly damped oscillations with a frequency
proportional to the strength of an applied magnetic field. This opens up an
alternate way to sensitive magnetometric measurements. Applications in
low-light nonlinear optics such as photon entanglement are feasible.

PACS numbers: 42.50.Gy, 07.55.Ge
\end{abstract}

\bigskip

\bigskip

\bigskip

In the last years there has been growing interest in the study of nonlinear
Faraday rotation (NFR) \cite{1} in coherent atomic media. Both the regime of
weak fields, low atomic coherence and low optical density \cite{2,3} and the
regime of strong field, large long-lived coherence and optically dense
medium \cite{4} have been extensively investigated. If in the first case the
ultranarrow magneto-optic resonances were observed with relaxation rate of 2$%
\pi \times $1Hz \cite{2}, in high light-power regime the large atomic
coherence giving rise to electromagnetically induced transparency (EIT) \cite
{5} allows to enhance essentially the nonlinear Faraday signal in optically
thick medium. These experiments \cite{4,6}, as well as the theoretical
estimations \cite{7}, demonstrate the significant potential of both
techniques to prove substantially the sensitivity of NFR-based magnetometers
to small magnetic fields.

\qquad An important feature of coherent atomic medium is its extremely high
dispersion near sharp atomic resonance lines such that a large NFR occurs
even at very weak magnetic fields. The simplest system displaying this
feature is a three-level $\Lambda $-atom with ground-state Zeeman sublevels
interacting with the right (RCP) and left (LCP) circularly polarized
components of optical light in the presence of a static magnetic field. In
usual terms of induced-coherence treatment, the RCP and LCP components of
the light can be represented as playing the role of coupling and probe
fields with respect to each other. Having equal intensities they excite a
maximal Zeeman coherence in high-power regime. On the other hand, the steep
dispersion at probe frequency in more general case of strong coupling and
weak probe fields \cite{8,9} has been used earlier \cite{10} to detect small
magnetic fields by measuring the probe phase shift via optical
interferometry. These investigations were aimed at improvement of
sensitivity to magnetic fields compared to conventional optical pumping
magnetometers\cite{11,12}.

\qquad We study here, for the first time, the magneto-optical rotation in a
strongly driven coherent medium. The model we are considering is a
four-level system (Fig.1) where, as before, the RCP and LCP components of
weak probe light E$_{p}$\ \ of frequency $\omega _{p},$ tuned to the 
%TCIMACRO{\TEXTsymbol{\vert}}%
%BeginExpansion
\mbox{$\vert$}%
%EndExpansion
1%
%TCIMACRO{\TEXTsymbol{>}}%
%BeginExpansion
\mbox{$>$}%
%EndExpansion
$\rightarrow |$3%
%TCIMACRO{\TEXTsymbol{>} }%
%BeginExpansion
\mbox{$>$}%
%EndExpansion
and 
%TCIMACRO{\TEXTsymbol{\vert}}%
%BeginExpansion
\mbox{$\vert$}%
%EndExpansion
2%
%TCIMACRO{\TEXTsymbol{>}}%
%BeginExpansion
\mbox{$>$}%
%EndExpansion
$\rightarrow |$3%
%TCIMACRO{\TEXTsymbol{>} }%
%BeginExpansion
\mbox{$>$}%
%EndExpansion
transitions respectively, create a Zeeman coherence, while a strong coupling
field E$_{c}$ with frequency $\omega _{c}$ drives the atom on the auxiliary
transition 
%TCIMACRO{\TEXTsymbol{\vert}}%
%BeginExpansion
\mbox{$\vert$}%
%EndExpansion
0%
%TCIMACRO{\TEXTsymbol{>}}%
%BeginExpansion
\mbox{$>$}%
%EndExpansion
$\rightarrow |$3%
%TCIMACRO{\TEXTsymbol{>}}%
%BeginExpansion
\mbox{$>$}%
%EndExpansion
. It is obvious, that in Faraday geometry the Zeeman sublevels 
%TCIMACRO{\TEXTsymbol{\vert}}%
%BeginExpansion
\mbox{$\vert$}%
%EndExpansion
1,2%
%TCIMACRO{\TEXTsymbol{>} }%
%BeginExpansion
\mbox{$>$}%
%EndExpansion
and the state 
%TCIMACRO{\TEXTsymbol{\vert}}%
%BeginExpansion
\mbox{$\vert$}%
%EndExpansion
0%
%TCIMACRO{\TEXTsymbol{>} }%
%BeginExpansion
\mbox{$>$}%
%EndExpansion
belong to different hyperfine components of the ground state. Moreover, the
coupling field E$_{c}$ should be circularly polarized. Our motivation for
the present work is related to the novel features of Faraday rotation
arising in this system in the EIT-regime. We show here that the Faraday
rotation angle oscillates in time with a slowly damped amplitude and a
frequency proportional to strength of magnetic field. For weak probe light
and small magnetic fields the damping rate of the oscillations is determined
by the lifetime of Zeeman coherence, which can be made very long compared to
the decay time of excited state . This opens up a promising possibility to
detect the small magnetic fields via the measurements of Faraday signal
oscillations with a sensitivity not less than already reached one.

\qquad The underlying mechanism of the long-lived oscillations of Faraday
signal is an existence of stable coherent superposition of two ground-state
sublevels 
%TCIMACRO{\TEXTsymbol{\vert}}%
%BeginExpansion
\mbox{$\vert$}%
%EndExpansion
1%
%TCIMACRO{\TEXTsymbol{>} }%
%BeginExpansion
\mbox{$>$}%
%EndExpansion
and 
%TCIMACRO{\TEXTsymbol{\vert}}%
%BeginExpansion
\mbox{$\vert$}%
%EndExpansion
2%
%TCIMACRO{\TEXTsymbol{>} }%
%BeginExpansion
\mbox{$>$}%
%EndExpansion
with different energies $\delta _{1,2}=\mp \delta ,$where $\delta $=g$\mu
_{B}$B/$\hbar $ \ is the Zeeman level shift induced by a magnetic field B (g
is a Lande factor). In the absence of magnetic field the strong coupling
field drives the atoms into population-trapped (or noncoupled) states $%
(|i\rangle -\frac{\Omega _{p}^{{}}}{\Omega _{c}^{{}}}|0\rangle )\symbol{126}$
$|i\rangle $, i=1,2$\ $($\Omega _{c}$ and $\Omega _{p}$ are the Rabi
frequencies of the coupling and probe fields: $\Omega _{c}>>\Omega _{p}$),
which are mixed with each other due to two-photon interaction with the
circular components of the probe light. This occurs at the time scale of
lifetime of upper state 
%TCIMACRO{\TEXTsymbol{\vert}}%
%BeginExpansion
\mbox{$\vert$}%
%EndExpansion
3%
%TCIMACRO{\TEXTsymbol{>}}%
%BeginExpansion
\mbox{$>$}%
%EndExpansion
. For all time thereafter the atom is in a steady-state and does not
interact with the fields. When a weak magnetic field is applied, the system
evolves in the same scenario with except that now the states 
%TCIMACRO{\TEXTsymbol{\vert}}%
%BeginExpansion
\mbox{$\vert$}%
%EndExpansion
1%
%TCIMACRO{\TEXTsymbol{>} }%
%BeginExpansion
\mbox{$>$}%
%EndExpansion
and 
%TCIMACRO{\TEXTsymbol{\vert}}%
%BeginExpansion
\mbox{$\vert$}%
%EndExpansion
2%
%TCIMACRO{\TEXTsymbol{>} }%
%BeginExpansion
\mbox{$>$}%
%EndExpansion
acquire different phases due to their different energies. This means that in
the EIT-regime the coherent superposition of the states $|1\rangle \exp
(-i\delta _{1}t)\ \ $and $|2\rangle \exp (-i\delta _{2}t)$ is eventually
created with the constant probabilities except for a very slow decay due to
Zeeman coherence damping. These superposition states exhibit the quantum
beats in time evolution of the system, which are revealed in NFR in the form
of slowly damped temporal oscillations. Thus, the key point is that to a
good approximation the time behavior of Zeeman coherence is the oscillatory
free evolution, but at the same time the susceptibility of probe transition,
although small, is different from zero that makes it possible to detect
these oscillations in Faraday signal.

\qquad We now turn to the analytical calculations of this effect in the
scheme depicted in Fig.1. The total interaction-picture Hamiltonian
corresponding to this system has the form

\begin{equation}
H=\hbar \Delta _{1}|1\rangle \langle 1|+\hbar \Delta _{2}|2\rangle \langle
2|+\hbar \Delta _{c}|3\rangle \langle 3|-\hbar (\Omega _{c}|3\rangle \langle
0|+\Omega _{1}|3\rangle \langle 1|+\Omega _{2}|3\rangle \langle 2|+H..c.)
\label{1}
\end{equation}
where the Rabi frequencies of coupling field and polarization components of
probe light are defined as $\Omega _{c}=\mu _{30}E_{c}/\hbar $ and $\Omega
_{1,2}=\Omega _{p}exp(i\Phi _{1,2}),\ \Omega _{p}=\mu E_{p}/\hbar ,$ $\mu
_{ik}$ is the dipole moment of the i-k transition, $\mu =\mu _{31}=\mu
_{32},\ $the amplitudes E$_{c,p}$\ are considered real and $\Phi _{1,2}\ $%
are the phases of probe polarization components. The detunings

\begin{equation}
\ \Delta _{c}=\omega _{30}-\omega _{c}-\delta _{0},\quad \Delta _{i}=\omega
_{3i}-\omega _{p}-\delta _{i},\quad i=1,2,
\end{equation}
include the frequency shifts of atomic transitions $\delta _{1,2}$\ and $%
\delta _{0}=g_{0}\mu _{B}Bm_{0}/\hbar \ $caused by a magnetic field, where g$%
_{0}$ and m$_{0}$ are the giromagnetic factor and magnetic quantum number
(momentum projection) of the level 
%TCIMACRO{\TEXTsymbol{\vert}}%
%BeginExpansion
\mbox{$\vert$}%
%EndExpansion
0$\rangle $. The thermal motion of the atoms with velocity $v$ in the
propagation direction is taken into account by incorporating into the
detunings the Doppler shifts $\omega _{c,p}v/c$, which are practically the
same for closely spaced hyperfine components of the ground atomic state.

\qquad The time evolution of the system's density matrix obeys the equation 
\begin{equation}
%TCIMACRO{\dfrac{d}{dt}}%
%BeginExpansion
{\displaystyle{d \over dt}}%
%EndExpansion
\rho =-%
%TCIMACRO{\dfrac{i}{\hbar }}%
%BeginExpansion
{\displaystyle{i \over \hbar }}%
%EndExpansion
[H,\rho ]+L\rho \ 
\end{equation}
\ where the operator L accounts for all relaxation processes. In our model
they are determined by the spontaneous emission from the upper level 
%TCIMACRO{\TEXTsymbol{\vert}}%
%BeginExpansion
\mbox{$\vert$}%
%EndExpansion
3$\rangle $ to the lower levels and the damping of lower-level coherence.
For a simplicity, all optical decay rates are assumed to be the same: $%
\gamma _{3i}$=$\gamma $, and the coherence decay rates of the lower levels
are taken equal $\gamma _{ij}$=$\gamma _{0}$%
%TCIMACRO{\TEXTsymbol{<}}%
%BeginExpansion
\mbox{$<$}%
%EndExpansion
%TCIMACRO{\TEXTsymbol{<}}%
%BeginExpansion
\mbox{$<$}%
%EndExpansion
$\gamma $, i, j=0,1,2.

\qquad In general case the Eq.(3) cannot be solved exactly. However, an
analytical solution can be found in symmetric case, when the coupling field
is tuned to resonance with 
%TCIMACRO{\TEXTsymbol{\vert}}%
%BeginExpansion
\mbox{$\vert$}%
%EndExpansion
0%
%TCIMACRO{\TEXTsymbol{>}}%
%BeginExpansion
\mbox{$>$}%
%EndExpansion
$\rightarrow |$3%
%TCIMACRO{\TEXTsymbol{>} }%
%BeginExpansion
\mbox{$>$}%
%EndExpansion
transition in the presence of magnetic field, i.e.$\Delta _{c}=0$, while for
the probe field the condition $\omega _{3i}-\omega _{p}=0,$ i=1,2 holds, so
that$\ \Delta _{1,2}=-\delta _{1,2}=\pm \delta .$ We consider the case of
strong coupling laser and weak magnetic fields such that the EIT conditions
are satisfied: 
\begin{equation}
\Omega _{c}^{2}>>\ \delta \Delta _{D},\ \gamma _{0}(\gamma +\Delta _{D}),
\end{equation}
\ where $\Delta _{D}$ is the Doppler width of the optical transitions. Then,
with the assumptions of \ $\Omega _{p},\delta <<\Omega _{c},\gamma ,\ $we
are able to derive the solution for the Zeeman coherence and the absorption $%
\alpha _{1,2}$ and dispersion $\beta _{1,2}$ coefficients of probe field in
the lowest order of small parameters $\Omega _{p}$/$\Omega _{c}$ and $\delta 
$/$\Omega _{c}$. With equal initial populations in the Zeeman sublevels $%
\rho _{11}$(0) =$\rho _{22}$(0)=0.5 this solution for time t much larger
than the upper level inverse decay rate: t%
%TCIMACRO{\TEXTsymbol{>}}%
%BeginExpansion
\mbox{$>$}%
%EndExpansion
%TCIMACRO{\TEXTsymbol{>}}%
%BeginExpansion
\mbox{$>$}%
%EndExpansion
$\gamma ^{-1},\ $has the form: 
\begin{equation}
\rho _{21}=-%
%TCIMACRO{\dfrac{_{{}}\Omega _{1}\Omega _{2}^{\ast }}{2\Omega _{c}^{2}}}%
%BeginExpansion
{\displaystyle{_{{}}\Omega _{1}\Omega _{2}^{\ast } \over 2\Omega _{c}^{2}}}%
%EndExpansion
\{1+\exp (-\gamma _{d}t)\exp [i(\ \delta _{1}-\delta _{2})t]\}\quad
+0(e^{-\gamma t})
\end{equation}
\begin{equation}
\alpha _{i}(t)=%
%TCIMACRO{\dfrac{1}{l_{0}}}%
%BeginExpansion
{\displaystyle{1 \over l_{0}}}%
%EndExpansion
%TCIMACRO{\func{Im}}%
%BeginExpansion
\mathop{\rm Im}%
%EndExpansion
(%
%TCIMACRO{\dfrac{\rho _{3i}\Gamma }{\Omega _{i}}}%
%BeginExpansion
{\displaystyle{\rho _{3i}\Gamma  \over \Omega _{i}}}%
%EndExpansion
),\quad \ \beta _{i}(t)=%
%TCIMACRO{\dfrac{1}{2l_{0}}}%
%BeginExpansion
{\displaystyle{1 \over 2l_{0}}}%
%EndExpansion
%TCIMACRO{\func{Re}}%
%BeginExpansion
\mathop{\rm Re}%
%EndExpansion
(%
%TCIMACRO{\dfrac{\rho _{3i}\Gamma }{\Omega _{i}}}%
%BeginExpansion
{\displaystyle{\rho _{3i}\Gamma  \over \Omega _{i}}}%
%EndExpansion
),\quad i=1,2
\end{equation}
$\ $where 
\begin{equation}
\rho _{31}=%
%TCIMACRO{\dfrac{_{{}}\Omega _{1}\delta _{1}^{{}}}{2\Omega _{c}^{4}}}%
%BeginExpansion
{\displaystyle{_{{}}\Omega _{1}\delta _{1}^{{}} \over 2\Omega _{c}^{4}}}%
%EndExpansion
(\Omega _{c}^{2}+i\Gamma \delta _{1})-%
%TCIMACRO{
%\dfrac{_{{}}\Omega _{1}|\Omega _{2}|^{2}\delta _{2}}{2\Omega _{c}^{4}}}%
%BeginExpansion
{\displaystyle{_{{}}\Omega _{1}|\Omega _{2}|^{2}\delta _{2} \over 2\Omega _{c}^{4}}}%
%EndExpansion
\exp (-\gamma _{d}t)\exp [i(\ \delta _{1}-\delta _{2})t]\ \ +0(e^{-\gamma t})
\end{equation}
\begin{equation}
\rho _{32}=\rho _{31}(1\longleftrightarrow 2),
\end{equation}
\begin{equation}
\gamma _{d}=\gamma _{0}+2\Gamma \Omega _{p}^{2}\delta ^{2}/\Omega _{c}^{4}
\end{equation}
\ \ Here $l_{0}=%
%TCIMACRO{\dfrac{1}{\sigma N}}%
%BeginExpansion
{\displaystyle{1 \over \sigma N}}%
%EndExpansion
,\ \sigma =4\pi \omega _{p}\mu ^{2}/\hbar c\Gamma $\ is resonant absorption
cross-section, $\Gamma =3\gamma /2$ is optical dephasing linewidth and N is
the atomic density. From the symmetry of the system with respect to two
circular components of probe light it follows that \ $\alpha _{1}(t)=\alpha
_{2}(t)=\alpha (t)$ and $\beta _{1}(t)=-\beta _{2}(t)$, as is apparent from
Eqs.(6-8). Note also that \ due to the condition $\Omega _{p}<<\Gamma \ $%
nearly all of the atomic population remains in the states 
%TCIMACRO{\TEXTsymbol{\vert}}%
%BeginExpansion
\mbox{$\vert$}%
%EndExpansion
1%
%TCIMACRO{\TEXTsymbol{>} }%
%BeginExpansion
\mbox{$>$}%
%EndExpansion
and 
%TCIMACRO{\TEXTsymbol{\vert}}%
%BeginExpansion
\mbox{$\vert$}%
%EndExpansion
2%
%TCIMACRO{\TEXTsymbol{>}}%
%BeginExpansion
\mbox{$>$}%
%EndExpansion
, so that\ the depletion of coupling field can be neglected.

\qquad The first terms in Eqs.(5) and (7,8) represent the results of EIT
regime in $\Lambda $-subsystems (0-3-1) and (0-3-2) and, of course, coincide
with that obtained in phase-sensitive model \cite{9} up to the factor 1/2%
{\sl \ }emerged from initial conditions. The immediate interaction between
the two polarization components of probe field in three-level configuration
(1-3-2) is responsible for the second terms, which reflect the quantum beats
in the system. It is worth noting that in the limit B$\rightarrow $0 the two
ways yield the same contribution to Zeeman coherence $\rho _{21}$. We
recognize from Eqs.(5-9) that in this limit there is no oscillation in the
system. At the same time it is seen that the frequency of the oscillations
is linear in the magnetic field, while their damping rate $\gamma _{d}\ $has
square dependence on B. As it follows from Eq.(9), for $2\Gamma \Omega
_{p}^{2}\delta ^{2}/\Omega _{c}^{4}$%
%TCIMACRO{\TEXTsymbol{<}}%
%BeginExpansion
\mbox{$<$}%
%EndExpansion
%TCIMACRO{\TEXTsymbol{<}}%
%BeginExpansion
\mbox{$<$}%
%EndExpansion
$\gamma _{0}$ the damping of oscillations is determined solely by Zeeman
decoherence rate $\gamma _{0}$. In Fig.2 we show the time evolution of the
real part of $\rho _{21}$ obtained by numerical integration of Eq.(3) for
two different values of magnetic field with the initial conditions $\rho
_{11}$(0) =$\rho _{22}$(0)=0.5 and the other $\rho _{ij}$=0 (i,j=0-3). As
expected, the oscillations occur in time region t%
%TCIMACRO{\TEXTsymbol{>}}%
%BeginExpansion
\mbox{$>$}%
%EndExpansion
%TCIMACRO{\TEXTsymbol{>}}%
%BeginExpansion
\mbox{$>$}%
%EndExpansion
$\Gamma ^{-1}$. Note that they always start with same value Re($\rho _{21}$)%
\symbol{126}$-|\Omega _{1}\Omega _{2}|/(\Omega _{c}^{2}+2\Omega _{p}^{2})$,
but according to Eq.(9) the amplitude of the oscillations is damped fast as
the strength of magnetic field increases. In the absence of magnetic field $%
\rho _{21}$\ is purely real and reaches very quickly the staedy-state value,
which is two times larger ( when disregarding the sign) than time averaged
value of Re($\rho _{21}$) at B$\neq 0$ (see Eq.(5)). It is not difficult to
be convinced, that the analytical solution Eq.(5) is represented graphically
by the same curves with insignificant deviation.{\sl \ }For comparison we
plot in Fig.2 the Zeeman coherence in the case of $\Delta _{c}=-\delta _{0},$
when the symmetry between the two circular components of probe field is
violated. However, we observe that this has a negligible effect, if magnetic
field is small. This analysis allows us to conclude that for sub-Gauss
magnetic fields our analytical results represent the quantitative physics
quite well. \ \ 

\qquad The equations of motion for the propagation of the intensity I$%
_{p}\thicksim \Omega _{p}^{2}$ and phases $\Phi _{1,2}\ $of the RCP and LCP
components of probe field read 
\begin{equation}
(%
%TCIMACRO{\dfrac{d}{dz}}%
%BeginExpansion
{\displaystyle{d \over dz}}%
%EndExpansion
+%
%TCIMACRO{\dfrac{1}{v_{g}}}%
%BeginExpansion
{\displaystyle{1 \over v_{g}}}%
%EndExpansion
%TCIMACRO{\dfrac{d}{dt}}%
%BeginExpansion
{\displaystyle{d \over dt}}%
%EndExpansion
)\ \Omega _{p}^{2}=-\ \alpha (t)\ \Omega _{p}^{2}\ ,
\end{equation}
\begin{equation}
(%
%TCIMACRO{\dfrac{d}{dz}}%
%BeginExpansion
{\displaystyle{d \over dz}}%
%EndExpansion
+%
%TCIMACRO{\dfrac{1}{v_{g}}}%
%BeginExpansion
{\displaystyle{1 \over v_{g}}}%
%EndExpansion
%TCIMACRO{\dfrac{d}{dt}}%
%BeginExpansion
{\displaystyle{d \over dt}}%
%EndExpansion
)\ \Phi _{1,2}^{{}}=\mp \beta _{2}(t),
\end{equation}
\ where the group velocities $v_{g}$\ expressed as 
\begin{equation}
v_{g}=c[1+c%
%TCIMACRO{\dfrac{\partial \beta _{i}}{\partial \delta _{i}}}%
%BeginExpansion
{\displaystyle{\partial \beta _{i} \over \partial \delta _{i}}}%
%EndExpansion
]^{-1}\approx 4\Omega _{c}^{2}l_{0}\Gamma ^{-1}
\end{equation}
are equal for two polarization components.

\qquad\ The most transparent results are obtained in the case of optically
thin media, where $\alpha L<<1,$L \ is the atomic cell length. Using the
Eqs.(6) and (7) this condition can be written as \ \ \quad\ 

\begin{equation}
\frac{\delta ^{2}\Gamma ^{2}L}{2\Omega _{c}^{4}l_{0}}<<1  \label{13}
\end{equation}

In this case the integration of Eqs.(10) and (11) yields for the probe
transmitted power and rotation angle $\Phi =(\Phi _{2}-\Phi _{1)}/2\ $%
\begin{equation}
\Omega _{p}^{2}(L,t)=\Omega _{p}^{2}(0,t-L/v_{g})\ =\Omega _{p}^{2}(0)
\end{equation}
\begin{equation}
\Phi (t)=\int\limits_{0}^{L}\beta _{2}(t-z/v_{g})dz=%
%TCIMACRO{\dfrac{\delta L}{v_{g}}}%
%BeginExpansion
{\displaystyle{\delta L \over v_{g}}}%
%EndExpansion
+%
%TCIMACRO{\dfrac{\Omega _{p}^{2}(0)}{\Omega _{c}^{2}}}%
%BeginExpansion
{\displaystyle{\Omega _{p}^{2}(0) \over \Omega _{c}^{2}}}%
%EndExpansion
\sin (%
%TCIMACRO{\dfrac{\delta L}{v_{g}}}%
%BeginExpansion
{\displaystyle{\delta L \over v_{g}}}%
%EndExpansion
)\cos (2\delta t)\exp (-\ \gamma _{d}t)
\end{equation}
where $\Omega _{p}^{{}}(0)$ corresponds to the input probe field. Here we
have used the Eq.(12) for $v_{g}$ and\ have taken into account that $\gamma
_{d}<<\delta .$

\qquad The Eq.(15) shows an interesting variation of oscillating term on
parameter\ $\eta =\delta L/v_{g}$, which is the ratio of probe propogation
time \ $L/v_{g}$\ to the atomic oscillation characteristic time \ $\delta
^{-1}$. This modulation can be understood rather symply. When $\eta <<1,$\
the atoms on propogation line are in the same phase of ground-state
oscillation during all passing of probe field through the medium. This leads
to total accumulation of polarization rotation and to the maximal relative
contribution of oscillating term in output rotation angle. In opposite case
of $\eta >>1$ the probe pulse moves very slowly and sees the atoms in
different phases of oscillation. As a result, the polarization rotation
varies periodically with distance travelled that eventually leads to the
sinusoidal dependence of the total rotation angle on the cell length. In
this case, of course, the relative contribution of oscillating term is
reduced, even if $\sin \eta =1$. By the same reasons, one can conclude that
in high density case ( $\alpha L>>1)\ $\ the oscillating term in $\Phi (t)\ $%
is suppressed even more.

\qquad\ The estimations with the help of Eq.(15) show that in the case of D$%
_{2}$ line in Rb the polarization rotation as high as 0.1 rad can be
achieved for sub-Gauss magnetic fields (B$\simeq $0.7G$)$, atomic density N%
\symbol{126}10$^{10}$cm$^{-3}$, L=3cm and $\Omega _{c}=\Gamma $ ($\alpha $L%
\symbol{126}0.1) with the amplitude of oscillations $\Omega _{p}^{2}$/$%
\Omega _{c}^{2}$\symbol{126}0.1. For large values of $\eta ,$\ we should
make sure of compatibility of the conditions $\eta >>1\ $and $\alpha L<<1$.
Combining the parameters $\alpha ,\eta \ $and using the expression of $v_{g}$%
\ from Eq.(12) the latter condition can be written in the form $2\eta \delta
\Gamma /\Omega _{c}^{2}<<1,\ $which can be fulfilled in a wide range of $%
\eta $, if one takes into account that $\delta <<\Omega _{c}^{{}}$.The
predicted effect is much easy to be observed in a cooled atomic ensemble,
where the Doppler broadening can be neglected that allows to avoid the
strict restrictions of Eq.(4) on the magnetic field.

It is worth noting that the present technique can be used to fully entangle
a pair of very weak optical fields. There have been several studies\cite
{13,14}predicting an unprecedental strength of nonlinear interaction between
two low-power light pulses. The attempts to apply these schemes to
generation of photon entangled states\cite{15} have confronted with the
problem of the mismatch between the slow group velocities of propagating
weak fields that restricts their effective interaction time\cite{16}. To
overcome this difficulty it was proposed in \cite{15} to use the mixture of
different isotopes of alkali atoms and to apply an appropriate magnetic
field. To demonstrate how the efficient interaction of two weak pulses
arises in our scheme we consider the case of two uncorrelated weak beams
with equal frequencies instead of two circular components of linearly
polarized probe light\ and suppose that the frequency of the fields is
chosen such that, say, the first field is in exact resonance with the 
%TCIMACRO{\TEXTsymbol{\vert}}%
%BeginExpansion
\mbox{$\vert$}%
%EndExpansion
1%
%TCIMACRO{\TEXTsymbol{>}}%
%BeginExpansion
\mbox{$>$}%
%EndExpansion
$\rightarrow |$3%
%TCIMACRO{\TEXTsymbol{>} }%
%BeginExpansion
\mbox{$>$}%
%EndExpansion
transition in the presence of magnetic field, i.e. $\Delta _{1}=0,$\ while $%
\Delta _{2}=2\delta .\ $Then, when putting in Eq.(7) $\delta _{1}=0$\ and\ $%
\delta _{2}=2\delta $, we find that the phase shift and absorption of the
first field are proportional to the intensity of the second one that creates
the cross-modulation between weak beams. The main advantage of our scheme is
that the problem of group velocity mathching does not arise here, since, as
it was shown above, the two beams have the same group velocities and, hence,
the interaction between them is maintained for a very long time.

In summary, we have described new phenomenon of long-lived temporal
oscillations of NFR, which arise when the conventional Faraday configuration
is driven by a strong laser field, and have shown that these oscillations
reflect the quantum beats in coherent superposition of ground-state Zeeman
sublevels. We have found the analytical solution for total angle of Faraday
rotation,which allows to understand quantitatively the main features of the
effect. Our study predicts also the ability of the proposed scheme to
achieve an efficient creation of two-photon entanglement. This question will
be discussed in detail elsewhere.

\bigskip

{\bf Figure captions}

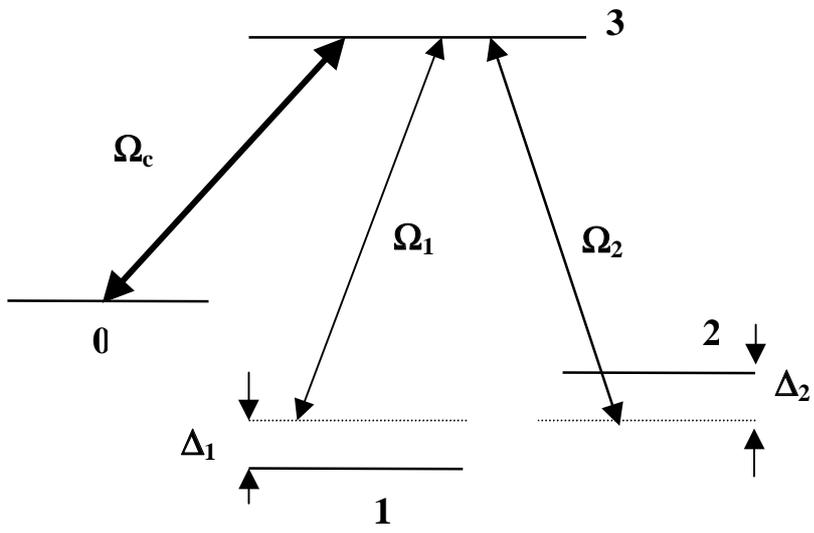
FIG.1. Four-level model for observation of oscillating Faraday rotation. $%
\Omega _{c}$ and $\Omega _{1,2}$ are the Rabi frequencies of coupling field
and $\sigma _{\pm }\ $components of probe light, respectively. $\Delta
_{2,1} $ are the shifts of \ m=$\pm 1$ \ levels induced by magnetic field.

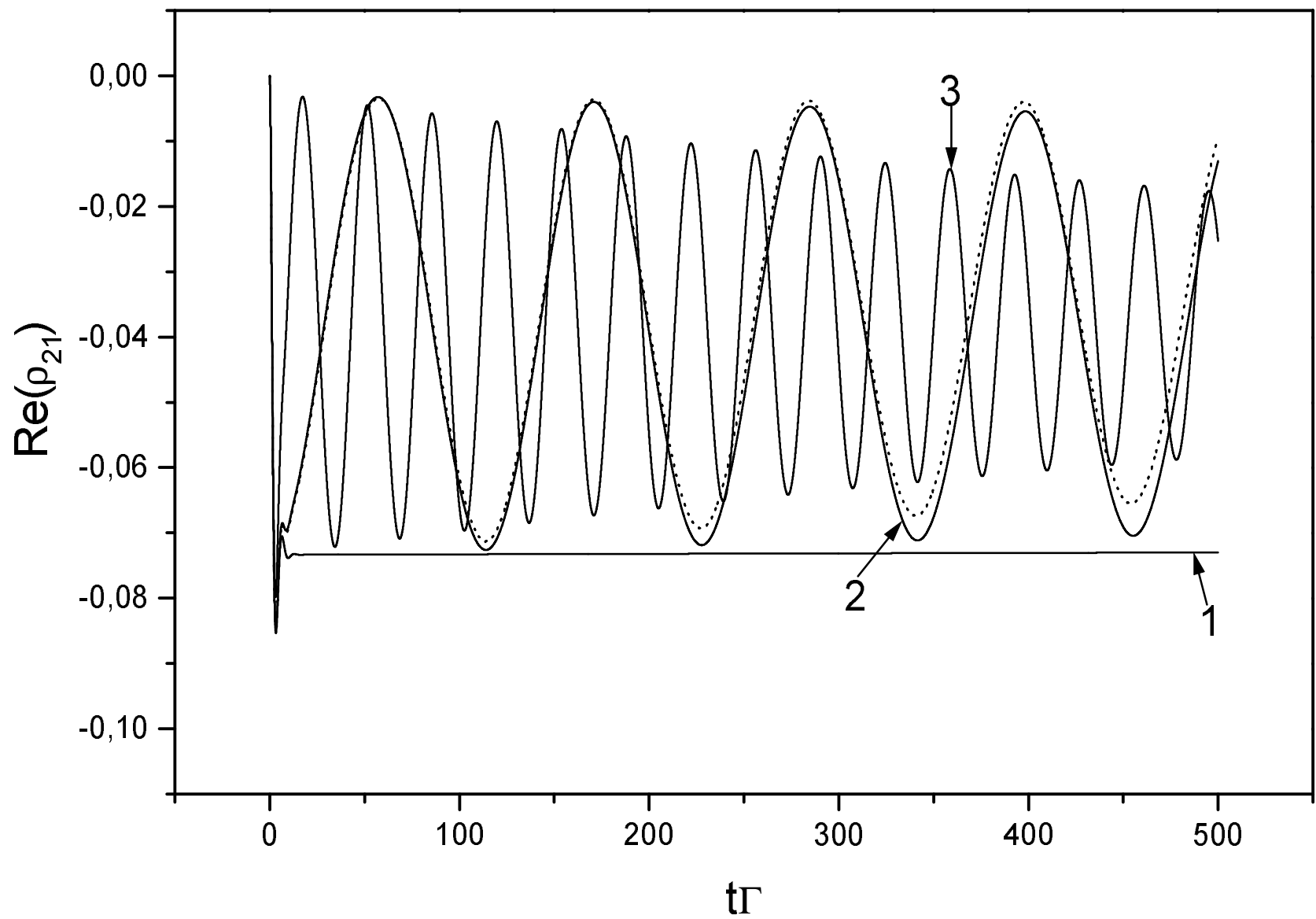
FIG.2. Numerical results for time evolution of Zeeman coherence Re($\rho
_{21}$) in the case of $\Omega _{c}$=$\Gamma $, $\Omega _{1,2}$=0.3$\Gamma $%
, $\gamma _{0}$=10$^{-4}\Gamma ,\ \Delta _{c}=0$ and for different values of
magnetic field B: 1) $\delta =0,\ 2)$ $\delta $=0.03$\Gamma $ and 2) $\delta 
$=0.1$\Gamma $. The{\sl \ }dotted line corresponds to the case of \ $\Delta
_{c}=\delta $=0.03$\Gamma $ with the same values of the rest parameters. The
Doppler broadening is ignored.

\end{document}